\documentclass[twocolumn,aps,prl,epsf,showpacs]{revtex4}

\usepackage{amsmath} 
\usepackage{amsfonts} 
\usepackage{epsfig} 
\usepackage{epsf} 
\usepackage{array}

\begin{document}

\title{Enhanced Superconductivity in Superlattices of High-$T_c$ Cuprates} 
\author{Satoshi Okamoto$^1$}
\author{Thomas A. Maier$^2$}
\affiliation{$^1$Materials Science and Technology Division, Oak Ridge National Laboratory, Oak Ridge, Tennessee 37831-6071, USA \\
$^2$Computer Science and Mathematics Division and 
Center for Nanophase Materials Sciences, Oak Ridge National Laboratory, Oak Ridge, Tennessee 37831-6164, USA}

\begin{abstract}
The electronic properties of multilayers of strongly correlated models for cuprate superconductors are investigated 
using cluster dynamical mean-field techniques. 
We focus on combinations of underdoped and overdoped layers and find that the superconducting order parameter in the overdoped layers 
is enhanced by the proximity effect of the strong pairing scale originating from the underdoped layers. 
The enhanced order parameter can even exceed the maximum value in uniform systems. 
This behavior is well reproduced in slave-boson mean-field calculations which also find higher transition temperatures than in the uniform system. 
\end{abstract}

\pacs{71.10.Fd, 74.20.-z, 74.78.Fk} 
\maketitle

The fabrication and characterization of artificial heterostructures involving metallic-ion oxides 
is one of the main topics of current materials science \cite{Izumi01,Ohtomo02,Okamoto04a}. 
Heterostructure might be useful to explore new and better functionalities that cannot be achieved in the bulk form. 
Among various oxides, of particular interest are correlated-electron materials \cite{Imada98} 
which are known to exhibit rich phenomena including high-$T_c$ superconductivity in the cuprates \cite{Bednorz96}. 

In this Letter, we are interested in the properties of model superlattices (SL's) for high-$T_c$ cuprates with different doping levels, 
in particular, a combination of underdoped (UD) and overdoped (OD) cuprates. 
In the former, a high pairing scale is observed both experimentally \cite{Gomes07} and theoretically \cite{Maier06}, 
while in the latter, stronger coherence of carriers exists, although $T_c$ is lower than at optimal doping in both the regions. 
By connecting the two doping regions in a multilayer geometry, the proximity of the strong pairing interaction from the UD region 
may produce stronger superconductivity in the OD region. 
Similar proposals are seen in Refs.~\cite{Hayashi03,Ginzburg05}. 
Conversely, the phase stiffness of a superconductor with small superfluid density may be increased by coupling it to a good metal 
resulting in a higher $T_c$ \cite{Berg08}.
The effects of spatial inhomogeneity in (quasi-)two-dimensional systems have been investigated in the light of the nanoscale inhomogeneity 
often observed in UD cuprates \cite{Kivelson05,Tsai06,Tsai08,Doluweera08}. 
In contrast to such in-plane inhomogeneity, the decoherence effect is small in SL's. 
Further, various conditions of the SL can be controlled by present experimental capabilities. 
In fact, enhanced superconductivity in such systems was reported recently \cite{Yuli08,Smadici08,Gozar08}.

Here, we provide further insight into this problem by studying Hubbard and $t$-$J$ models of SL's made up of UD and OD cuprate layers 
using layer extensions of the cellular dynamical mean-field theory (CDMFT) \cite{Kotliar01,Kotliar06} 
and the dynamic cluster approximation (DCA) \cite{Hettler:dca,Maier05} as well as the slave-boson mean-field (SBMF) theory.
We find that the superconducting (SC) order parameter is indeed increased in UD/OD SL's as compared to its value in the uniform system. 
For some combinations of dopings, the SC order parameter is found to become larger than its largest value in the uniform system. 
Furthermore, the SBMF calculation predicts a transition temperature in the SL that exceeds the maximum $T_c$ in the uniform system. 
These results suggest that ``higher-$T_c$'' superconductivity may indeed be realized in SL's of cuprates. 

We first consider a quasi-two-dimensional (Q2D) Hubbard model for the Cu $d$ electrons: 
\begin{eqnarray}
H_{Hub} \!\!&=&\!\! \sum_{i, \sigma} \varepsilon (z_i) n_{i \sigma} + U \sum_i n_{i \uparrow} n_{i \downarrow} 
- t \hspace{-1em} \sum_{\langle ij \rangle \in xy, \sigma} \hspace{-0.8em} \bigl(c_{i \sigma}^\dag c_{j \sigma} + H.c. \bigr) \nonumber \\
&& -t_z \hspace{-0.8em} \sum_{\langle ij \rangle \in z, \sigma} \hspace{-0.6em} \bigl(c_{i \sigma}^\dag c_{j \sigma} + H.c. \bigr) . 
\label{eq:Hhub} 
\end{eqnarray}
Here, $c_{i \sigma}$ is the annihilation operator of an electron with spin $\sigma$ at site $i$ and $n_{i \sigma} = c_{i \sigma}^\dag c_{i \sigma}$, 
$t (t_z)$ the nearest neighbor transfer integral along the $xy (z)$ directions, $U$ the on-site repulsive Coulomb interaction, 
and $\varepsilon (z_i)$ the on-site potential dependent on the layer index $z_i$. 
In general, the potential $\varepsilon (z_i)$ should be computed self-consistently by including the long-ranged Coulomb interaction 
between electrons and between electrons and the background charge. 
Instead of performing such a fully self-consistent calculation, 
we fix the potential profile as well as the Fermi level by focusing on short-period SL's which have a small number of inequivalent layers. 
Thus, the charge density profile should be regarded as a result of the charge transfer among the constituent layers \cite{Ohtomo02,Okamoto04a}. 

To begin, we study the zero temperature SC order parameters of SL and uniform systems using the CDMFT.  
We treat the interlayer correlations on the mean-field level \cite{Okamoto04b} and take care of short-ranged intralayer correlations 
by using a square $2 \times 2$ cluster. 
In this approximation, the self-energy 
depends on the layer index $z_i$ and the intralayer cluster coordinates $\vec R_i$. 
The CDMFT then maps the bulk lattice problem onto a number $N_z$ of independent effective cluster problems, 
with $N_z$ the number of layers in the unit cell. 
The cluster problems are solved by using the Lanczos exact diagonalization (ED) technique \cite{Caffarel94,Kancharla05}.
We use $U = 10 t$ and $t_z =0.5 t$. Results do not depend on the choice of these parameters in a significant way 
unless $U \ll W$ (the total band width) 
or $t_z=0$ (purely two-dimensional case) or $t_z =t$ (three dimensional case). 

\begin{figure}[tbp] 
\includegraphics[width=0.8\columnwidth,clip]{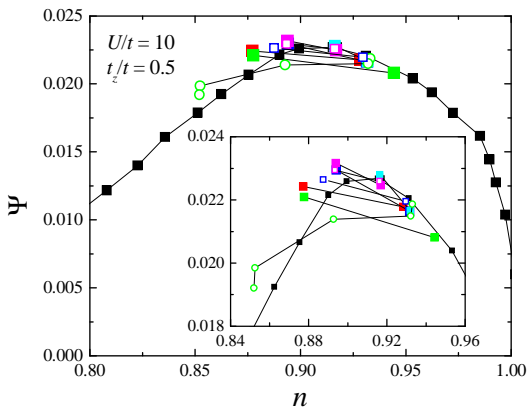} 
\caption{(Color online) The superconducting order parameter $\Psi$ as a function of carrier density $n$ for the Q2D Hubbard model computed 
by CDMFT with the ED impurity solver at $T=0$. Black symbols are the results for the uniform systems. 
Colored (shaded) symbols connected by a thin line denote the results for the SL. The inset shows the magnification around the optimally doped region. } 
\label{fig:PsiED} 
\end{figure}

Results for the $d$-wave SC order parameter 
$\Psi = \sum_\sigma \langle \sigma c_{i \sigma} c_{i+x \bar \sigma} \rangle = - \sum_\sigma \langle \sigma c_{i \sigma} c_{i+y \bar \sigma} \rangle$ 
for the uniform system are shown as black dots in Fig.~\ref{fig:PsiED}. 
A ``dome'' like shape is consistent with previous calculations for the two-dimensional Hubbard model \cite{Kancharla05}. 
We find an optimal carrier density $n^{opt}=0.91$ which maximizes $\Psi$. 

Next, we construct a SL in which UD layers with $n \agt n^{opt}$ and OD layers with $n \alt n^{opt}$ are repeated along the $z$ direction. 
We plot $\Psi (z_i)$ at the corresponding densities $n=n(z_i)$ in Fig.~\ref{fig:PsiED} as colored (shaded) squares connected by thin lines. 
Filled (open) squares are the results for SL's in which two (one) UD layers and two (one) OD layers are repeated, 
i.e. AABB (ABAB) superlattices. 
One sees that $\Psi$ in the OD layers is enhanced. 
The enhancement is stronger in AABB superlattices than in ABAB superlattces 
possibly due to the dilution in the latter. 

We also considered thicker SL's consisting of several UD layers and OD layers sandwiching nearly optimal-doped layers, e.g. AA'BCC' superlattices. 
Even in this case, we found the enhancement in the OD layers. 
However, $\Psi$ in the nearly optimal-doped layers is suppressed because the pairing interaction is reduced by the OD layers, 
and we never observed significantly larger order parameters than in the uniform system. An example is shown by open circles in Fig.~\ref{fig:PsiED}. 

To gain more insight into the mechanism responsible for the enhanced superconductivity in the SL and obtain estimates for $T_c$, 
we next apply a SBMF approximation to a layer $t$-$J$ model, 
the large $U$ limit of the Hubbard model in which no double occupancy is allowed and nearest neighbor spins are coupled by exchange interactions $J$ and $J_z$. 
The Hamiltonian is given by 
\begin{eqnarray}
H_{tJ} \!\!&=&\!\! \sum_{i, \sigma} \varepsilon(z_i) \tilde n_{i \sigma} - t \hspace{-1em} \sum_{\langle ij \rangle \in xy, \sigma} 
\hspace{-0.8em} \bigl( \tilde c_{i \sigma}^\dag \tilde c_{j \sigma} + H.c. \bigr) + J \hspace{-0.5em} \sum_{\langle ij \rangle \in xy} 
\hspace{-0.5em} \vec S_i \cdot \vec S_j \nonumber \\
&& -t_z \hspace{-0.8em} \sum_{\langle ij \rangle \in z, \sigma} \hspace{-0.6em} \bigl(\tilde c_{i \sigma}^\dag \tilde c_{j \sigma} + H.c. \bigr) 
+ J_z \hspace{-0.5em} \sum_{\langle ij \rangle \in z} \hspace{-0.5em} \vec S_i \cdot \vec S_j, 
\label{eq:Htj} 
\end{eqnarray}
where, $\vec S_i = \sum_{\sigma \sigma'} \tilde c_{i \sigma}^\dag \vec \tau \tilde c_{i \sigma'}$ with $\vec \tau$ the Pauli matrices, 
$\tilde c_{i \sigma}$ is a projected fermion operator defined as $\tilde c_{i \sigma}=c_{i\sigma}(1-n_{i \bar \sigma})$, 
and $\tilde n_{i \sigma} = \tilde c_{i \sigma}^\dag \tilde c_{i \sigma}$. 
Here, we take $t_z=0.5t$, $J = 0.4 t$ and $J_z=0$ for simplicity. 
The qualitative behavior does not depend on the choice of exchange interactions. 

In the SBMF approximation one introduces fermionic spinon $f_\sigma$ and bosonic holon $b$ operators 
with a local constraint $b_i^\dag b_i + \sum_\sigma f_{i \sigma}^\dag f_{i \sigma} = 1$ and rewrites the electron operators as 
$\tilde c_{i \sigma} = b_i^\dag f_{i \sigma}$. 
By introducing a mean-field decoupling in $H_{tJ}$; 
$\chi^f = \sum_\sigma \langle f_{i \sigma}^\dag f_{i+x \sigma} \rangle = \sum_\sigma \langle f_{i \sigma}^\dag f_{i+y \sigma} \rangle$ 
[uniform resonance valence bond (RVB)], 
$\chi^f_z = \sum_\sigma \langle f_{i \sigma}^\dag f_{i+z \sigma} \rangle$, 
$\Delta = \sum_\sigma \langle \sigma f_{i \sigma} f_{i+x \bar \sigma} \rangle = - \sum_\sigma \langle \sigma f_{i \sigma} f_{i+y \bar \sigma} \rangle$ 
(singlet RVB), 
$\chi^b = \langle b_{i}^\dag b_{i+x} \rangle = \langle b_{i}^\dag b_{i+y} \rangle$, $\chi^b_z= \langle b_{i}^\dag b_{i+z} \rangle$, 
and relaxing the local constraint to a global one, we obtain the mean-field Hamiltonians for spinons and holons \cite{Baskaran87,Kotliar88,Affleck88}. 

Solving the self-consistent equations is straightforward. 
However, we find that 
the onset temperatures of $\chi^b_z$ (and therefore $\chi^f_z$) and holon condensation $N_0 = \langle b_i \rangle^2$ almost coincide 
for a relatively wide range of $t_z/t$. 
Therefore, such a treatment is not useful to address the effect of the interlayer coupling on the SC transition. 
In fact, these phase transitions are artifacts of the mean-field decoupling. 
To avoid these fictitious phase transitions, we replace $\chi^f$ and $\chi^f_z$ (weakly dependent on $n$ and $T$ at $n \sim 1$) 
in the holon Hamiltonian with a small number $r$ of the order of 0.1, 
and perform a self-consistent calculation for the other part. 
This is similar to approximating $\chi^b = \delta = 1-n$ \cite{Baskaran87}. 
But, the present treatment can be used in a wider parameter range since the $n$ and $T$ dependences of $\chi^b$ and $\chi^b_z$ are included. 
There remain phase transitions associated with $\Delta$ and $N_0$. 
A higher transition temperature between them is expected to become a smooth crossover due to the gauge fluctuations, 
and only the lower temperature, i.e. SC critical temperature $T_c$, remains as a true phase transition \cite{Nagaosa90}. 
Since our main interest is in obtaining an estimate of $T_c$, we keep our discussion on the mean-field level. 

\begin{figure}[tbp] 
\includegraphics[width=0.8\columnwidth,clip]{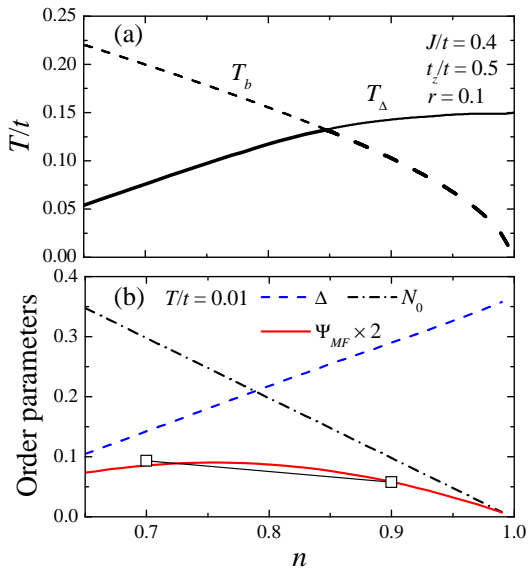} 
\caption{(Color online) (a) Mean-field phase diagram of the Q2D $t$-$J$ model as a function of carrier concentration $n$ and temperature $T$ 
with parameters as indicated. 
Below $T_b$ holon bose condensation $N_0$ occurs, while below $T_\Delta$ the singlet RVB order parameter $\Delta$ becomes finite. 
(b) Order parameters as a function of $n$ at $T = 0.01 t$. 
Broken line: $\Delta$, dash-dotted line: $N_0$, 
and solid line: $d$-wave superconducting order parameter $\Psi_{MF}$. 
The squares indicate the order parameters $\Psi_{MF}$ of a SL in which layers with $n=0.7$ and 0.9 alternate. } 
\label{fig:PD} 
\end{figure}

\begin{figure}[tbp] 
\includegraphics[width=0.8\columnwidth,clip]{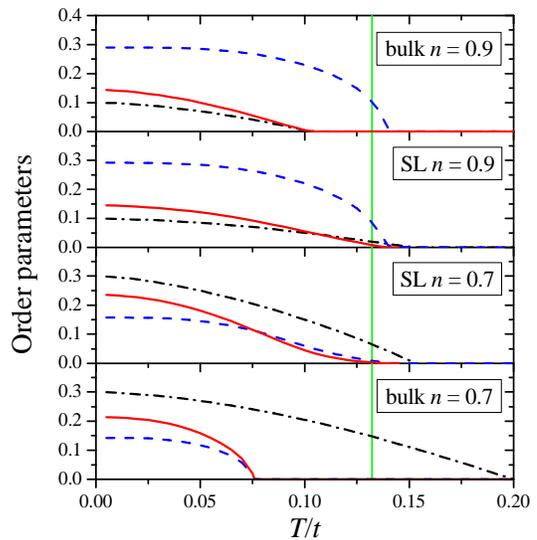} 
\caption{(Color online) Mean-field order parameters of the Q2D $t$-$J$ model as a function of $T$ with the same parameters as in Fig.~\ref{fig:PD}. 
Top (bottom): Uniform system with $n=0.9 (0.7)$. 
Middle: Layers with $n=0.9$ and 0.7 in the SL. 
Broken lines: $\Delta$, dash-dotted lines: $N_0$, and solid lines: $\Psi_{MF} (\times 5)$. 
Vertical line indicates $T_c^{opt} = 0.132 t$ of the uniform system. } 
\label{fig:orderparameter} 
\end{figure}

Numerical result for the phase diagram is shown in Fig.~\ref{fig:PD}~(a) as a function of carrier concentration $n$ and temperature $T$. 
Below $T_b$, holon condensation occurs, and below $T_\Delta$, $\Delta$ becomes finite. 
Thus, the lower temperature signals the onset of $d$-wave superconductivity. 
Optimal doping is located at $n^{opt}=0.846$ with a maximum critical temperature $T_c^{opt} = 0.132 t$. 
The characteristic $T$ dependence of the order parameters of the uniform system at under doping ($n=0.9$) and over doping ($n=0.7$) 
are displayed in the top panel and the bottom panel of Fig.~\ref{fig:orderparameter}, respectively. 
In Fig.~\ref{fig:PD}~(b), we show the order parameters at low temperature as a function of $n$. 
In the UD region strong singlet correlations exist but weak coherence, while in the OD region this behavior is reversed. 
Consequently, the mean-field SC order parameter $\Psi_{MF} = N_0 \Delta$ has a maximum ($\Psi_{MF}^{max} = 0.0451$) as $T_c$ 
but at a slightly smaller carrier density $n =0.76$. 

Next, we consider a SL in which UD ($n=0.9$) and OD ($n=0.7$) layers alternate along the $z$ direction (ABAB stacking for simplicity). 
At low enough temperature, most holons are condensed giving rise to $N_0\sim 0.1 (0.3)$ in the UD (OD) layers. 
Since there are enough carriers (holons) to be paired in the OD layers, 
the proximity of $\Delta$ from the UD layers enhances the SC order parameter in the OD layer ($\Psi_{MF} = 0.0467 > \Psi_{MF}^{max}$) 
as shown by the squares in Fig.~\ref{fig:PD} (b). 
On the other hand, in the UD layers, most carriers are already paired by strong pairing interactions. 
Thus, the interlayer coupling does not affect the UD layers significantly. 
These results are consistent with the CDMFT results for Hubbard SL's in Fig.~\ref{fig:PsiED}.  

The temperature dependence of the SL order parameters is shown in the middle two panels of Fig.~\ref{fig:orderparameter}. 
The onset of $\Delta$ is dominated by the UD layer which has a higher $T_\Delta$ in the uniform system, so $T_\Delta = 0.14t$. 
The onset of $N_0$ is roughly determined by the average carrier density $n^{av}=0.8$ and $T_b \sim 0.152 t$ 
because it represents three-dimensional coherence. 
Thus, the lower temperature corresponds to $T_c = T_\Delta = 0.14t$, which exceeds the maximum $T_c$ in the uniform system, $T_c^{opt} = 0.132 t $, 
by about 6\%. 
When the carrier density in the OD layer is further reduced, $T_b$ is increased but $T_\Delta$ is decreased. 
Therefore, an optimal combination is naturally expected. 

Finally, we check the finite temperature SBMF results using a layer extension of the DCA. 
The DCA for the SL also maps the bulk system onto a number of independent effective cluster models embedded in a dynamic host. 
In contrast to the CDMFT, the cluster is defined in reciprocal space. 
We solve the cluster problems with a noncrossing approximation (NCA) which has been shown to provide qualitatively similar results 
to a numerically exact quantum Monte Carlo method \cite{Maier00,Maier05}.

\begin{figure}[tbp] 
\includegraphics[width=0.9\columnwidth,clip]{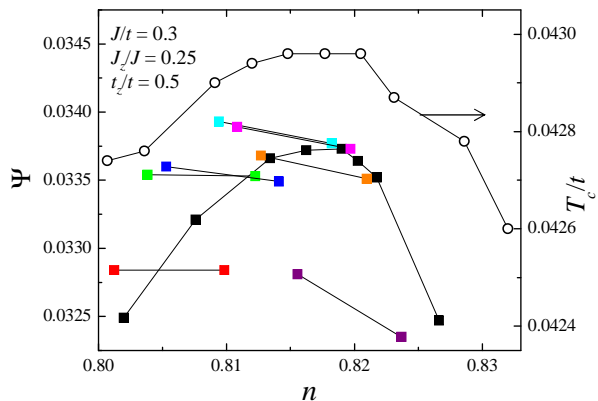} 
\caption{(Color online) 
The superconducting order parameter $\Psi$ as a function of carrier density $n$ for the Q2D $t$-$J$ model computed by the DCA-NCA 
at a temperature $T = 0.041 t$ close to $T_c$. 
The results for the uniform systems are indicated by the black symbols, while the colored (shaded) symbols connected by a thin line denote the result for the SL. 
Also shown is doping dependence of $T_c$ in the uniform system (open circles). } 
\label{fig:PsiDCA} 
\end{figure}

As can be seen from Fig.~\ref{fig:orderparameter}, the order parameter in the SL system is very small close to $T_c$. 
The same behavior is observed in the DCA-NCA calculations for the SL, which renders a reliable extraction of $T_c$ difficult if not impossible. 
We therefore present DCA-NCA results for the order parameter at a temperature below the $T_c$ of the uniform and SL system 
and argue that an enhancement of the order parameter in the SL close to $T_c$ should translate to an increase in $T_c$ over the uniform system.

Figure~\ref{fig:PsiDCA} summarizes the results of the layer DCA calculations. 
Here we used $t_z=0.5t$, $J=0.3t$, and an additional interlayer exchange $J_z=0.25J$. 
In the uniform system we find an optimal doping $n^{opt} \approx 0.82$ with $T_c^{opt} \approx 0.043t$. 
When rescaled by $J$, this $T_c^{opt}$ is about a factor 2.4 lower than that in the SBMF calculation. 
The order parameter $\Psi$ at a temperature $T=0.041t$ slightly below $T_c$ shows a similar $n$ dependence as $T_c$. 
For the AABB SL system, we again find a clear enhancement of the order parameter $\Psi$ in the OD layers which can even exceed the optimal value in the uniform system. 
This behavior is consistent with the SBMF calculations at finite $T$ and the CDMFT at $T=0$. 

Calculations for $t_z < 0.5t$ (not shown) indicate that both the order parameter and $T_c$ are reduced, 
and, as a result, the enhancement of superconductivity. 
Note that in the limit of $t_z=0$, the enhancement has to vanish. 

The present mean-field calculations suggest a ``recipe'' to increase $T_c$ in SL's of cuprates: 
(1) Combine UD cuprates with large pairing scale and OD cuprates with strong coherence and 
(2) make the average carrier density $n^{av}$ slightly smaller than the bulk optimal value $n^{opt}$. 
The latter condition is required to obtain a larger order parameter, which gives rise to higher $T_c$ even in the double-layer systems 
when there is enough phase stiffness. 
Requiring strong coherence, i.e., smaller $n^{av}$ than $n^{opt}$, is related to the proposal in Ref.~\cite{Berg08}. 
A definite proof that this mechanism can give rise to higher $T_c$ requires calculations with fully three-dimensional clusters. 
Results of these calculations will be published elsewhere.

To summarize, we studied the properties of model superlattices consisting of underdoped and overdoped high-$T_c$ cuprates. 
Two cluster dynamical-mean-field methods were extended for the superlattice geometry and applied to Hubbard and $t$-$J$ models 
for the cuprate superconducting superlattices and compared to a more transparent slave-boson mean-field calculation. 
All calculations show enhanced superconductivity in the overdoped layers resulting from the proximity of strong pairing correlations 
from the underdoped layers. 
At certain combinations of doping, the superconducting order parameter even exceeds the largest value in the uniform systems. 
The slave-boson mean-field study shows a higher superconducting transition temperature than the maximum $T_c$ in the uniform system. 
%

%
The authors thank H. Seo, M. Hayashi, and D. J. Scalapino for their discussions. 
S. O. acknowledges kind suggestions by B. Kyung, A.-M. S. Tremblay, and M. Eisenbach for developing a numerical code. 
This work was supported by the Division of Materials Sciences and Engineering, Office of Basic Energy Sciences, U.S. Department of Energy. 
A portion of this research at Oak Ridge National Laboratory's Center for Nanophase Materials Sciences was sponsored 
by the Scientific User Facilities Division, Office of Basic Energy Sciences, U.S. Department of Energy.


\end{document}